# A MapReduce-based rotation forest classifier for epileptic seizure prediction

Samed Jukic
International Burch University, Faculty of Engineering and Information Technologies, Francuske Revolucije bb. Ilidza, Sarajevo, 71000, Bosnia and Herzegovina.
E-mail: samed.jukic@ibu.edu.ba;

Abdulhamit Subasi
Effat University, College of Engineering, Jeddah, 21478, Saudi Arabia
E-mail: absubasi@effatuniversity.edu.sa

*Abstract*—In this era, big data applications including biomedical are becoming attractive as the data generation and storage is increased in the last years. The big data processing to extract knowledge becomes challenging since the data mining techniques are not adapted to the new requirements. In this study, we analyse the EEG signals for epileptic seizure detection in the big data scenario using Rotation Forest classifier. Specifically, MSPCA is used for denoising, WPD is used for feature extraction and Rotation Forest is used for classification in a MapReduce framework to correctly predict the epileptic seizure. This paper presents a MapReduce-based distributed ensemble algorithm for epileptic seizure prediction and trains a Rotation Forest on each dataset in parallel using a cluster of computers. The results of MapReduce based Rotation Forest show that the proposed framework reduces the training time significantly while accomplishing a high level of performance in classifications.

## 1. Introduction

Epilepsy is a neurological disorder characterized by a frequent tendency of the brain to yield abrupt bursts of abnormal electrical activity [1]. Such occurrences are called seizures and occur randomly. Excessive and synchronized activity of neurons causes epileptic seizures [2, 3]. Epilepsy is the second most common neurological disorder after strokes affecting over 1% of the world's population [4]. In order to diagnose and identify the epileptic seizures, mostly the patient's EEG must be monitored for several days. The process of monitoring is boring, time-consuming and expensive. Neurologist needs to determine the enduring epileptic activity from the recorded EEG data in order to determine if the used medication is working or not. Hence, a computer aided epileptic seizure detection system is extremely important [5, 6].

Electroencephalogram (EEG) has been used for clinical diagnosis of epilepsy for many decades. Compared to other methods such as Electrocorticogram (ECoG), EEG is a safe and clean method for detecting the activity of the brain. Clinical analysis of EEG traces for identification of seizures is well established. However, the performance of automated EEG based methods is dependent on the types of features analyzed and how they are used to classify the signal. Patients with epilepsy suffer from repeated seizures that manifest as physical or behavioral changes, which require intervention using medications or surgery [7]. A lot of channels are used for recording EEG signals. Processing of that number of channels is time consuming. Because of that, parallel processing is very important aspect of EEG signal processing in order to decrease processing time.

Since EEG signals with many channels and complicated signal processing and classification algorithms cannot be analysed easily by means of personal computers. The cloud computing is the practical solution to these kind of big data problems [8]. Hadoop is considered to work on cloud build from thousands of commodity hardware nodes. Biomedical signal processing and classification algorithms can be parallelized to save computing time. Message passing interface (MPI) which is a traditional parallelization method may cause whole procedure to fail. Hadoop is a fault tolerant platform which makes it natural choice for these types of algorithms. Hadoop has lately been utilized in different areas of big data analysis [9]. Parallel computing is a concurrent computing which uses a group of autonomous processors employed together to solve a large computational problem. The essence of parallel computing is to partition and distribute the whole computational work among the processors [10]. Google's MapReduce programming model [11] offers an effective structure for processing large datasets in a parallel manner. The Google File System [12] which inspires MapReduce delivers effective and consistent distributed data processing essential for applications including large datasets. The basic role of the MapReduce model is to support parallelism in which the programmer can benefit from the issues of distributed and parallel programming. Furthermore, MapReduce implementation includes load balancing, network performance, fault tolerance etc. [11]. The Apache Hadoop [13] is the most widely used open-source implementation of Google's MapReduce which is written in java for scalable, reliable distributed computing [14]. Hence, in this paper we used a MapReduce implementation of epileptic seizure prediction with wavelet packet decomposition (WPD) and Rotation Forest classifier in a Hadoop environment. The rest of the paper is organized as follows. The methodology and experimental design is given in Section 2. The results are presented and analysed in section 3. Discussion and conclusion is given in section 4 and 5 respectively.

## 2. Methodology and Experimental Design

The EEG data were collected at the Epilepsy Center of the University Hospital of Freiburg and visually inspected by expert neurologists. Recordings are sampled 256 Hz using 60 channels. EEG recordings are taken from 21 patients with 88 seizures including preictal data and seizure-free (interictal) [15].

All database signals are separated into ictal and interictal folders. Every seizure has preictal period for about one hour, so we used to combine preictal and ictal file and create 48 minutes of preictal EEG signal. For interictal signals 1 file defines 1 hour of EEG signal, and for preictal it depends on length of seizure. All patients have 24 or 25 hours, where we used 15 hours for training dataset, and rest 9 or 10 hours for test phase. For training dataset, generally, we used randomly chose chunk of 10 minutes, but for one patient we had to use all interictal signals for training dataset.

### 2.1. Multiscale Principal Component Analysis (MSPCA)

Principal Component Analysis combines the variables as a linear weighted sum transforms an $n \times p$ data matrix, X as

$$X = TP^T \quad (1)$$

where, n and p are the measurements and the principal component loadings notated as P and the principal component scores are defined as T. A significant decision in PCA is to choose the appropriate principal components that capture the essential relationship. Some methods are available for this task, and are studied by Jackson [16] and Malinowski [17]. A cross-validation can be used, if an approximation of the error is not available [18, 19, 20]. Multiscale PCA (MSPCA) is combination of the PCA ability to remove the cross-correlation between the variables. The observations are decomposed using wavelet transform for each variable to combine the PCA and benefits of wavelets. This results in data matrix transformation, X into a matrix, WX, where W is an $n \times p$ orthonormal matrix demonstrating the orthonormal wavelet transformation. The quantity of principal components to be reserved at each scale is not transformed by the wavelet decomposition because it doesn't change the fundamental relationship between the variables at any scale [18, 19, 21].

### 2.2. Discrete Wavelet Transformation (DWT) and Wavelet Packet Decomposition (WPD)

One of the effective time-frequency techniques for analysis of various non-stationary signals, such as EEG, belong to a group of wavelets-based methods. Discrete Wavelet Transform decomposes a signal $x(t)$ into a set of functions (wavelet coefficients) by scaling and shifting of mother wavelet function. The signal $x(t)$ can be rebuilt as linear combination of wavelets and weighting wavelet coefficients.

Procedure for DWT decomposition starts with the selection of the number of wavelet decomposition levels denoted as j_max. For the first decomposition level j = 1, discrete-time EEG signal, $x[k]$, is passed through the high-pass filter, $h[\cdot]$, and low-pass filter, $l[\cdot]$, and then downsampled by 2. The corresponding outputs are Detail, $D_j$, and Approximation, $A_j$ respectively:

$$D_j[i] = w_{high}[i] = \sum_k x[k] \cdot h[2 \cdot i - k] \quad (2)$$

$$A_j[i] = w_{low}[i] = \sum_k x[k] \cdot l[2 \cdot i - k] \quad (3)$$

After $D_j$ and $A_j$ have been obtained, the approximation $A_j$ is set as $x[k]$ and j is set as 2 (increased by 1), and the aforementioned procedure is repeated until j exceeds j_max [22].

Wavelet Packet Decomposition (WPD) is identical to DWT except that the detail coefficients $D_j$ are further decomposed as well. For a k-level wavelet decomposition, WPD will produce $2^k$ different sets of wavelet coefficients (each levels has its own approximation and detail record), whereas DWT generates $k+1$ sets of wavelet coefficients (each level has its own detail coefficient plus one final approximation).

### 2.3. Ensemble ML Techniques

Individual person usually cannot make as good decision as can make groups of people, particularly when each members of group come in with their own biases. In machine learning we have the same situation. Combination of multiple learners opinions construct learning models called Ensemble methods. On that way, we may usually try with applying many simpler learners and again get good performance. Furthermore, if you have multiple processors access, ensembles are inherently parallel, which may give better efficient at training and test time [23]. An ensemble of classifiers is a several classifiers collection whose discrete decisions are combined in some method to classify the test examples [24]. It is general thing that an ensemble usually gives much better results than the individual classifiers that make it up [25].

#### 2.3.1. Rotation Forest

Rotation Forest is another recently introduced effective ensemble classifier generation method [26], where the training set for every base classifier is made by using PCA to rotate the initial attribute axes. Precisely, to generate the training data for a base classifier, the attribute set F is randomly divided into K subsets and PCA is used to every subset. All principal components are kept because of preserving the variability data information. Therefore, K axis rotations are positioned to generate the new attributes for a base classifier. The key point of Rotation Forest is to simultaneously inspire diversity and individual accuracy inside the ensemble: diversity is presented while applying feature extraction for every base classifier and accuracy is required by storing all principal components and also using the entire data set to train each base classifier [27].

### 2.4. MapReduce and Hadoop

Big data applications mostly need more resources than available inexpensive machine [28]. The need for effective algorithms of parallel computing is apparent since the existence of extremely large datasets cannot be processed without using multiple computers. Google's MapReduce programming model [11] offers an effective framework for big data processing in a parallel manner. The Google File System [29] which motivates MapReduce delivers competent and consistent distributed computing for big data applications [30].

The MapReduce framework is used to iterate over the input, calculate the key/value pairs from each portion of the input, cluster all intermediate values by key, then iterate over the resulting groups and finally reduce each group. The model effectively provides parallelism. The basic role of the MapReduce model is to support parallelism in which the programmer can benefit from the issues of distributed and parallel programming. Furthermore, MapReduce employment includes network performance, load balancing, fault tolerance etc. [11]. The Apache Hadoop [13] is the widely used open-source application of Google's MapReduce for scalable, reliable and distributed programming in Java. Two different steps are employed in the application of the MapReduce model [28];

- Map: An initial transformation step, in which individual input records are processed in parallel.

- Reduce: A summarization step, in which all associated records are processed together by a single entity.

The Hadoop implementation of the MapReduce model is given in Fig. 1. It splits the input into logical chunks which is processed independently by a map task. The consequences of these processing chunks can be substantially divided into distinct sets, which are then sorted. Each sorted chunk is passed to a reduce task.

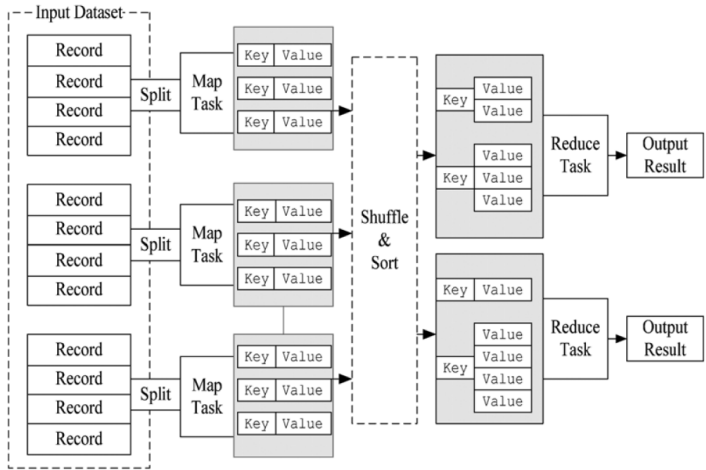

FIGURE 1 THE MAPREDUCE MODEL

The Hadoop File System (HDFS) is considered for MapReduce jobs which read input in large chunks, process it, and write chunks of output. File data is replicated to multiple storage nodes for reliability [28]. HDFS implemented by two processes:

- NameNode: Manages the file system metadata, and provides management and control services.

- DataNode: Provides block storage and retrieval services [30].

### 2.5. Parallel Processing

Parallel or concurrent computing denotes a group of autonomous processors working together to solve a computational problem. This needs to diminish the execution time and employ larger memory/storage resources. The use of parallel computing is to divide and distribute the whole computational task among the processors. But, the hardware architecture of any multi-processor system is rather different than a single-processor computer which requires specifically adapted parallel software [10].

MATLAB's Parallel Computing Toolbox is used to solve computationally and data-intensive problems employing multicore processors, GPUs, and computer clusters. High-level parallel for loops, special array types, and parallelized numerical algorithms allow to parallelize MATLAB applications without CUDA or MPI programming. You can use this toolbox with Simulink to run multiple simulations of a model in parallel. Furthermore, this toolbox can execute applications for the full processing power of multicore desktops with MATLAB computational engines running locally. You can run the same applications on a computer cluster or a grid computing service using MATLAB Distributed Computing Server™ without modifying the code [31].

In our experiment, we test three types of MATLAB computing: Running MATLAB code normally, multithreaded parallelism (MATLAB parallel) and explicit parallelism (Code parallel). One instance of MATLAB automatically creates multiple concurrent instruction streams in multithreaded parallelism. Multiple processors or cores, sharing the memory of a single computer, execute these streams. In explicit parallelism, numerous examples of MATLAB run on several processors or computers, mostly with distinct memories, and concurrently execute a single MATLAB command or M-function. New programming concepts, including parallel loops and distributed arrays, describe the parallelism [31].

### 2.6. Proposed Method

For experiment, we first check the number of seizures for each patient in Freiburg DB. Once we pick the patient with the most number of seizures, then we check the length of each seizures, and chose patient with longest seizures as it is presented in Table 1. After that we divide seizures into 2 groups according to their length. So, 3 longer seizures are going to train and the rest 2 are going to test. For every signal, we took 3 channels of ictal and preictal signals. We took those signals and generate new one with 48 minutes of preictal plus length of ictal. We check interictal signals for chosen patient and divide them into train and test (15 for train and 10 for test). 3 channels of each signal are loaded and create structure from them. After that we divide signals on 2048 chunks and generate new one with those parts. Three channels of interictal and preictal data from the training dataset will be divided into smaller segments (8 seconds – 2048 samples). The matrix will be made for every 8 minutes of these segments (the matrix size will be 2048 x 180). The matrix will be denoised using MSPCA. Wavelet packet decomposition features will be extracted from each denoised 8-seconds long segment and put into training database. 10-fold cross validation will be performed on the training database to tune the machine learning algorithm parameters. The window time of 8 minutes was chosen as we have 48 minutes of preictal time (contains 6 of these time windows). Five of these time windows (40 minutes) will be used for making any prediction of upcoming seizure. We will find as alarm if 3 continues chunks predicted as seizure from 5 chunks.

In explicit (code) paralleilization method we made parallel code in the part where it is possible. So we applied this paralleilzation in loading raw signals and in segmentation of signals where we made chunks of 8 seconds. This logic is used in signal processing for training dataset and in testing process. We could not apply explicit parallelization in part of code where processing order is important like de-noising process or feature extraction part. Beside parallelization method, we also applied Hadoop structure in signal processing part. We used five slaves and one master machine and distirubte job among all of them.

After we take interictal and ictal signals and create datatable from all signals together, then we applied RF machine learning to train our model. Our model is trained in MATLAB GUI application which is based on java library of WEKA and has option to save trained model. So after we train model, we save it and load in testing process. In test phase, we first load all interictal signals and one precital and ictal. Applied de-noising and feature extraction, and after that test data on trained model and see prediction value.

### 3. RESULTS

Unlike any other known seizure prediction method, our algorithm provides real-time prediction combining with parallel signal processing to achieve less time of execution. We mentioned before that our proposed method is patient oriented, which means that each patient has different machine learning trained model and result of every patient is different case.

For this experiment, we calculate number of seizure and seizure length for each patient, and then pick first five patient which have the biggest seizure length.

Table 1 represents accuracy which is achieved using training dataset for every of five patients that we used in this experiment. After we finished with training algorithm and get these accuracies, we save machine learning models and use them for testing.

TABLE 1 ACCURACY WHICH ACHIEVED FOR TRAINING DATASET

| Patient | Accuracy |
| --- | --- |
| Patient 3 | 90.45% |
| Patient 10 | 96.87% |
| Patient 11 | 99.87% |
| Patient 14 | 89.85% |
| Patient 16 | 94.53% |

Beside real-time prediction, in this experiment we introduce parallelization and Hadoop implementation for signal processing and using that we significantly reduce execution time, as you can see in Table 2, which represents average execution time. Execution time using code parallel method is double less then when we are using MATLAB parallel way of execution or normal - serial execution. When we apply code parallel method in Hadoop environment, we get significantly less execution time than single machine code parallel. This is especially important for testing case, because execution time is high, so applying code parallelization in Hadoop environment we get execution time more than 2 times lower than normal or MATLAB parallel. This difference is better representing in Figure 2.

TABLE 2. COMPARISON OF AVERAGE EXECUTION TIME IN SECONDS

| Patient | Normal execution | MATLAB parallel | Code parallel | Hadoop execution |
| --- | --- | --- | --- | --- |
| Patient3 | 215.08 | 211.05 | 117.7 | 92.39 |
| Patient10 | 247.91 | 245.52 | 124.93 | 98.76 |
| Patient11 | 228.6 | 223.88 | 111.77 | 88.36 |
| Patient14 | 232.61 | 229.63 | 123.02 | 95.78 |
| Patient16 | 233.09 | 231.84 | 119.28 | 90.1 |

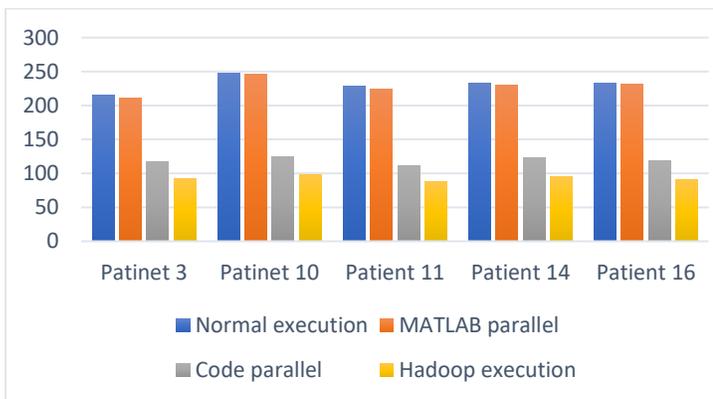

FIGURE 2 COMPARISON OF AVERAGE EXECUTION TIME IN SECONDS

In figure 3 is presented first seizure prediction for Patient 3 where blue bars are predicted values for each 8 minutes' chunk and red part of bars represent alarm state, position where patient will get notification that seizure is going to happened. For testing phase of patient 3 we used 9 hours of interictal signals and 1 hour of preictal and ictal signal. Every 1 hour is represented as 7.5 blue bars. As you can see, we did not predict preictal stage at the first time when it is occurred, but right after that we predict that there will be seizure, and system predict seizure around 30 minutes before it happened, which is great result.

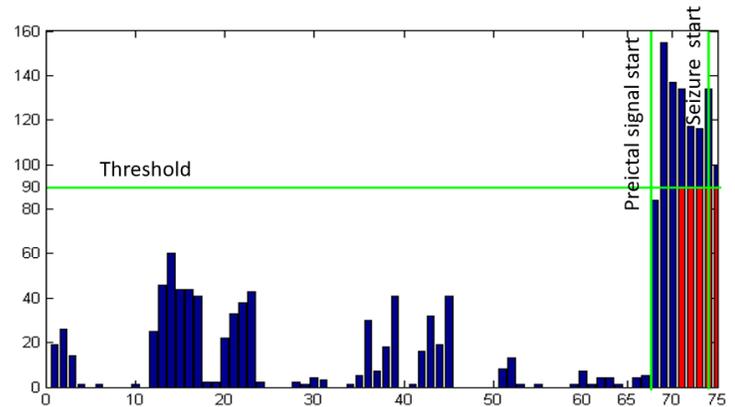

FIGURE 3 EPILEPTIC SEIZURE PREDICTION FOR FIRST SEIZURE OF PATIENT 3

In previous graph, we present prediction for first seizure, and in the following Figure 4 we introduce seizure prediction in combination of 9 hours of interictal signals, 1 hour of preictal and 1 hour of preictal and ictal signal. As you can see, we predict seizure 50 minutes before it happened.

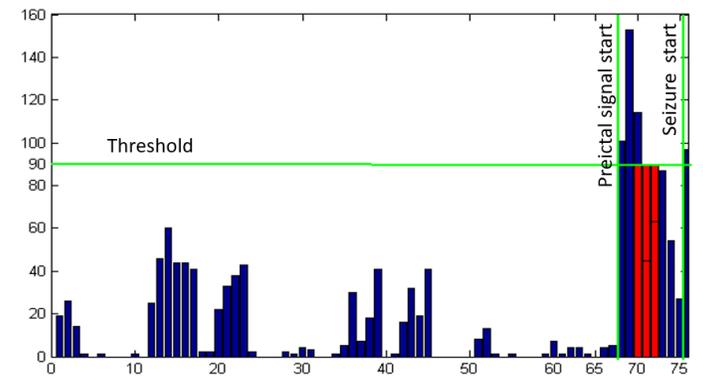

FIGURE 4 EPILEPTIC SEIZURE PREDICTION FOR SECOND SEIZURE OF PATIENT 3

Table 3 represents difference in execution time in each segment of processing signals. As we can see, in every segment, code parallel method and Hadoop execution is giving more than 2 times better results than Normal execution or MATLAB parallel way of execution.

TABLE 3 COMPARISON OF EXECUTION TIME IN SECONDS FOR PATIENT 3

| Patient 3 | Normal execution | MATLAB parallel | Code parallel | Hadoop execution |
| --- | --- | --- | --- | --- |
| Interictal signal processing | 105.14 | 101.98 | 62.53 | 48.23 |
| Preictal signal processing | 64.36 | 62.46 | 49.96 | 41.7243 |
| Signal processing total | 169.50 | 164.44 | 112.49 | 89.95 |
| Test | 521.32 | 515.34 | 245.82 | 189.6547 |

In Figure 5 is represent real-time seizure prediction for patient 10. We used 10 hours of interictal signals, 1 hour of preictal and 1 hour of preictal and ictal signal. As you can see, we predict seizure around 1 hour before seizure happened, and right after preictal signal start, we predict that as preictal phase. Besides that, there is also couple of situation where we predict chunk of 8 minutes as preictal phase even it was interictal, but it was there was not alarm, because it should be 3 out of 5 chunks predicted as preictal phase for alarm to be raised.

In Figure 6 we present seizure prediction for second seizure of Patient 10. We used the same structure like in previous case. As you can see, we predict seizure around 1 hour before it happened. Like in previous case, again we predict some interictal phase as preictal, but alarm did not appear because it should be predicted 3 out of 5 chunks as preictal.

Difference in execution time for patient 10 is as big as it is for patient 3 and you can see results in following Table 4. For test case, code parallel structure and Hadoop implementation is more than double faster than normal execution or MATLAB execution, which makes our system more reliable for implementation.

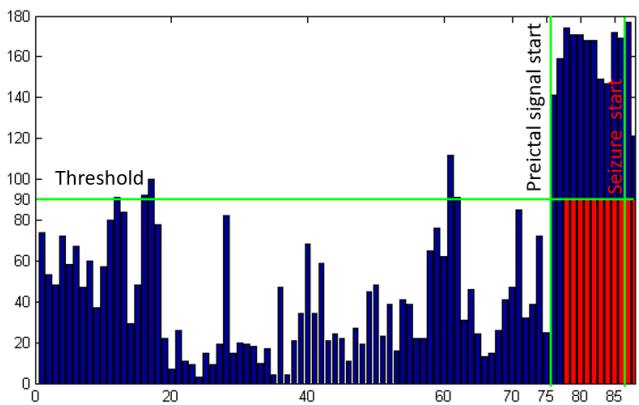

FIGURE 5 EPILEPTIC SEIZURE PREDICTION FOR FIRST SEIZURE OF PATIENT 10

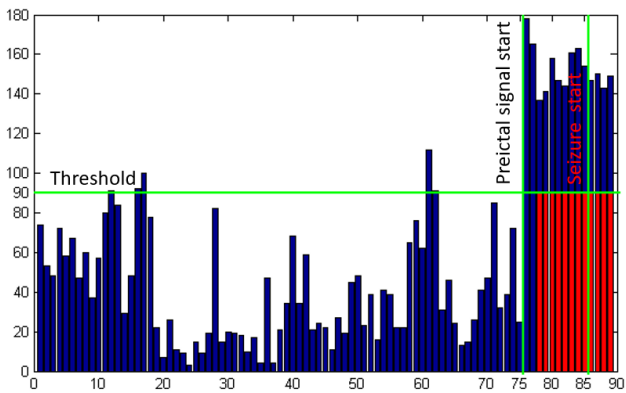

FIGURE 6 EPILEPTIC SEIZURE PREDICTION FOR SECOND SEIZURE OF PATIENT 10

TABLE 4 COMPARISON OF EXECUTION TIME IN SECONDS FOR PATIENT 10

| Patient 10 | Normal execution | MATLAB parallel | Code parallel | Hadoop execution |
|---|---|---|---|---|
| Interictal signal processing | 100.27 | 97.73 | 60.50 | 47.92 |
| Preictal signal processing | 65.94 | 64.20 | 51.90 | 43.21 |
| Signal processing total | 166.21 | 161.92 | 112.40 | 91.14 |
| Test | 659.23 | 658.25 | 274.92 | 212.79 |

For patient 11 we could not get sufficient result when we applied our normal way of construction training dataset, where we used only 10 minutes from each interictal signals, so for patient 11 we used all 1h of interictal signals and make much bigger dataset then we used for patient 3 and patient 10. In following Figure 7 you can see prediction accuracy. We raised alarm around 40 minutes before seizure happened which is great result. Besides that, as you can see that in interictal signals there is no any mistake in prediction which prove our assumption that bigger training dataset provides much better accuracy. For patient 11 we have only one seizure for testing.

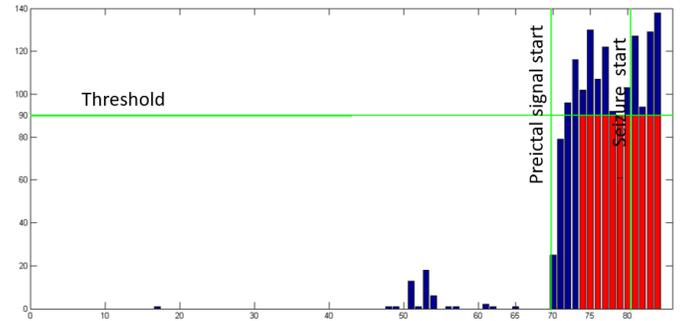

FIGURE 7 EPILEPTIC SEIZURE PREDICTION FOR PATIENT 11

Time execution table is represented in following Table 5 and difference in time execution is really huge if we compare normal execution and MATLAB parallel with code parallel method or Hadoop execution.

TABLE 5 COMPARISON OF EXECUTION TIME IN SECONDS FOR PATIENT 11

| Patient 11 | Normal execution | MATLAB parallel | Code parallel | Hadoop Execution |
|---|---|---|---|---|
| Interictal signal processing | 105.52 | 100.16 | 62.53 | 49.90 |
| Preictal signal processing | 36.86 | 36.45 | 30.91 | 23.65 |
| Signal processing total | 142.38 | 136.61 | 93.44 | 73.55 |
| Test | 629.65 | 622.30 | 260.18 | 206.33 |

For patient 11, we also compare accuracy of training dataset using 10-fold cross validation and results are represented in Table 6. We got that accuracy for 10 minutes of interictal signals is better than when we used 1 hour. But in testing cases, prediction algorithm which is created from dataset which includes 1 hour of interictal signals giving much better results.

TABLE 6 COMPARISON IN TRAINING ACCURACY FOR DIFFERENT LENGTH OF INTERICTAL SIGNALS FOR PATIENT 11

| Type | Accuracy |
|---|---|
| 10 min | 99.87 % |
| 1h | 97.31 % |

We applied proposed system for Patient 14, where we tried with 10 minutes and 1 h of interictal signals, but with both method we could not get any sufficient result. As you can see in Figure 8 accuracy for prediction is poor, and there is no any alarm raised in this test case.

Even if we did not get good result in real-time seizure prediction, we can state that parallel execution shows good results in terms of execution time. As you can see in Table 7, time which is need for execution of MATLAB code with explicit code parallel or Hadoop implementation is more than two times faster than two other proposed methods.

We also did comparison of accuracy in training dataset and get results which are presented in Table 8. As you can see, for patient 14 we got better results with bigger training dataset, even none of trained model did not give sufficient result in test case.

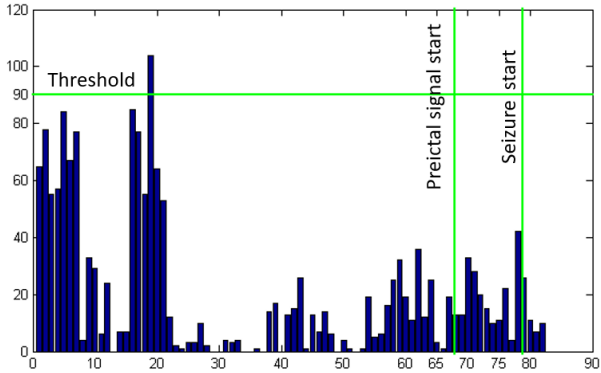

FIGURE 8 EPILEPTIC SEIZURE PREDICTION SEIZURE FOR PATIENT 14

TABLE 7 COMPARISON OF EXECUTION TIME IN SECONDS FOR PATIENT 14

| Patient 14 | Normal execution | MATLAB parallel | Code parallel | Hadoop Execution |
|---|---|---|---|---|
| Interictal signal processing | 102.59 | 101.02 | 66.86 | 52.01 |
| Preictal signal processing | 37.17 | 37.06 | 31.73 | 25.42 |
| Signal processing total | 139.75 | 138.08 | 98.59 | 77.43 |
| Test | 650.94 | 642.38 | 294.92 | 228.27 |

TABLE 8 COMPARISON IN TRAINING ACCURACY FOR DIFFERENT LENGTH OF INTERICTAL SIGNALS FOR PATIENT 14

| Type | Accuracy |
|---|---|
| 10 min | 89.85 % |
| 1h | 93.62 % |

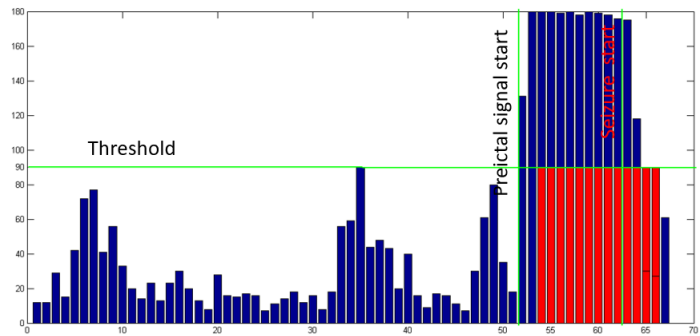

FIGURE 9 EPILEPTIC SEIZURE PREDICTION FOR FIRST SEIZURE OF PATIENT 16

Last patient that we made experiment is patient 16 and results of prediction accuracy for first seizure you can find in Figure 9. For this patient, we used 7 hours on interictal signals, 1 hour of preictal and 1 hour of preictal and ictal signal. As you can see from the figure, prediction accuracy is really high, whiteout any false prediction of preictal state in interictal phase. We predict epileptic seizure around 70 minutes before it happened. For training dataset of this patient, we used initially proposed method, where we used randomly taken 10 minutes from each interictal signal.

For second seizure prediction of patient 16 we used the same amount of interictal signals and different preictal and ictal signal. Accuracy of seizure prediction is represented in Figure 10. As you can see, we got really good accuracy and we predict seizure around 70 minutes before seizure starts.

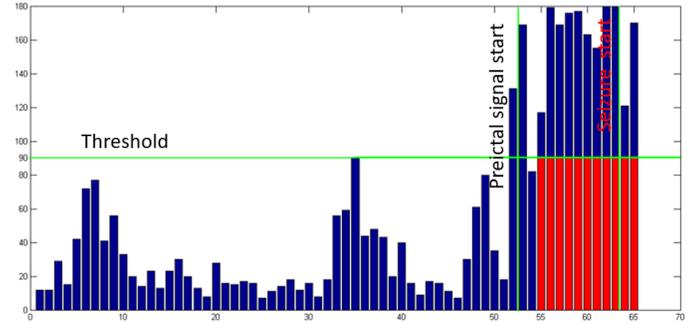

FIGURE 10 EPILEPTIC SEIZURE PREDICTION FOR SECOND SEIZURE OF PATIENT 16

Comparison of execution time for this patient is similar like in previous cases and it is represented in Table 9. As you can see, execution time for code parallel or Hadoop implementation is more than two times faster than normal or code parallel method. This difference is obvious for testing as much as it is for training phase.

TABLE 9 COMPARISON OF EXECUTION TIME IN SECONDS FOR PATIENT 16

| Patient 16 | Normal execution | MATLAB parallel | Code parallel | Hadoop Execution |
|---|---|---|---|---|
| Interictal signal processing | 101.80 | 101.24 | 62.12 | 48.14 |
| Preictal signal processing | 62.92 | 62.62 | 49.71 | 38.02 |
| Signal processing total | 164.71 | 163.86 | 111.82 | 86.16 |
| Test | 602.94 | 599.63 | 253.49 | 188.09 |

## 4. DISCUSSION

In this section, we discuss the findings and results of our proposed MapReduce based epileptic seizure prediction. If we look at our results, we can see that each patient is different story. First, we generate separate training dataset different for each patient, and based on that dataset we create trained model, then we load that model in for testing. Because we have different training datasets, we also have different model for each patient, and on that way, we increase the accuracy of seizure prediction. Our system is that much accurate also because we used majority in order to define alarm state. In our results, you can see couple of false prediction, especially for patient 10, but we did not mark that as alarm, because it happened only once or twice in bunch of five prediction. If it is happened that predict seizure 3 times in row, we count that as alarm state.

Another important thing related to our experiment is big difference in time execution of signal processing and testing when we used parallel processing. For most cases, speed of execution for code parallel way is two times faster than when we execute our code using ordinary way. Besides, we also compare time execution while we execute code on one machine one core, one machine multiple cores and multiple machine multiple cores (Hadoop cluster). Time which is needed to execute code on Hadoop cluster is much faster even when we execute code on multiple cores.

As we can see from Table 1, accuracy achieved on training dataset is not crucial in testing phase. Hence, for some patient we got really high training accuracy, but for testing it not achieve that good result, like Patient 11 and 14. Also, some for some patient we achieve around 90% accuracy for training, but it produces great accuracy in testing phase.

As we mentioned in the beginning, our system is patient dependent, we have different approach for every patient. Generally, we took randomly chosen 10 minutes of 1 hour from interictal signals and generate training dataset with those signal values. For patient 11, we tried with that logic, but we could not get good accuracy, and then we decide to take all interictal signals (1 hour) for training dataset. On that way we generate 6 times bigger training dataset, and based on that dataset, we create trained model with Rotation Forest machine learning method. Once we generate training dataset, we also test our Rotation Forest classifier on that dataset using cross-fold logic. With smaller dataset (10 minutes from each hour) we got better accuracy than when used all signals. It is interesting finding that overtraining gives lower results, but in testing phase, results of model which is generated from bigger dataset is much better.

For patient 14 we also tried with the same logic as patient 11, so first we tried with 10 minutes for training dataset, and accuracy was low, then we applied all signals, but again we could not get better results. For this patient, we again present decreasing in time execution using our method with code parallelization.

Our proposed method is completely real-time, because datasets that we are using for training and testing are completely separated. Therefore, for training phase we used 15 hours from interictal and 2 or 3 preictal and ictal phase, and with all of those signals we generate training dataset. For testing purpose, we used completely unknown signals, process them like we processed training signals, and send to trained model for evaluation. We are sending chunk of 8 minutes signal and make decision after every chunk. If 3 chunks out of 5 are categorized as seizure, then we define alarm state, and patient we get notification about that. Except for patient 14, for all other patients we made 100% accurate prediction, and inform patient about seizure at least half of hour before it happened.

In our experiment, we always used static threshold and it is half of total value for preictal signal, and compare our predicted value with threshold. It gives good accuracy, so we did not mention to change that way, but it will be better if we use dynamic threshold, which will be half of middle value for preictal predicted values.

If we check our results, you can see that we predict all seizures except seizures for patient 14, so we can say that for these 5 patients we have predict 7 out of 9 seizures which is 77.78% accuracy. But because our proposed system is patient oriented, we can say that we predict 100% accuracy for 4 patients and 0% for one patient.

Table 2 and Figure 1 represents comparison of average execution time for each patient. In results part we, also, present tables with separate time for signal processing and testing phase using different way of execution. In Table 2 we found average value for each of patient for every type of execution and compare that time. What is interesting to mention is that Normal execution (serial processing) and MATLAB parallel execution way is giving almost the same execution time. On the other hand, we have code parallel and Hadoop implementation as better methods which double reduce processing times. And if we check code parallel method and Hadoop implementation, we can state that Hadoop is for 20% or more faster than code parallel. This result is achieved because we used more machines with more cores in our experiment with Hadoop environment.

If we compare our work with other experiments, we could not find anyone who combines real-time and parallel or distributed way of processing, because of that we will to check our results with real-time prediction and distributed systems separately. In the study of Lian et. al. [32] they got accuracy 92%, seizure detection latency is 0.6 seconds, comparing with our experiment we got better accuracy for patient where we found our proposed method as suitable and our accuracy is 100% for prediction. Ali Shahidi et. al. [5] in their study using detection of epilepsy, and got accuracy around 90% latency of 7s comparing with our experiment where we are making seizure prediction with 100% accuracy. Beside these works, there is more experiment which are mentioned in literature review, and all of them are focusing on real-time detection and none of those papers are predicting epilepsy. Another group of papers are related to distributed systems and in research of Teixeiraa, et al., [33] they have 50% of predictions with false alarm of 1 in 6 h, comparing with our proposed system where we have 100% accuracy and no false alarm at all. In paper of Dutta et. al [34] they reduce time consuming and show that increase speed from half of hour processing to 12 minutes comparing with our experiment and parallelization where we increase speed 3 times. In the paper of Ježdík et. al. [35] they decrease time consuming for signal processing from 20 hours using one units to less than one hour when they used 18 units, where we also show that speed with more cores could be faster 3 or more times.

## 5. CONCLUSION

Nowadays, big data is getting attention as the huge amount of data are generated by different applications. The traditional data mining techniques can not handle the new requirements of big data. Therefore, we use the most popular MapReduce framework to deal with big amount of biomedical data. Besides, the Hadoop framework is the widely used open source implementation of MapReduce that enables the development of distributed solutions. In this paper, we have presented and evaluated a MapReduce based distributed Rotation Forest algorithm which exploits for automated epileptic seizure prediction. The performance of the proposed model is assessed in an experimental environment. The training dataset is divided into smaller subsets and the partitioned subsets are optimized across the cluster of multiple computing nodes. The proposed model reduces the training time significantly whereas a high level of accuracy is achieved in epileptic seizure prediction particularly for bulky training dataset.

ACKNOWLEDGMENT

The authors would like to thank to the University of Freiburg for providing the Freiburg Seizure Prediction EEG (FSPEEG) Database utilized in this research.